\documentclass{article}
\usepackage{amsmath,amsfonts}
\usepackage{algorithmic}
\usepackage{algorithm}
\usepackage{array}
\usepackage[caption=false,font=normalsize,labelfont=sf,textfont=sf]{subfig}
\usepackage{textcomp}
\usepackage{stfloats}
\usepackage{url}
\usepackage{verbatim}
\usepackage{graphicx}
\usepackage{cite}
\usepackage{amsthm}			%
\usepackage{amssymb}			%
\usepackage{algorithmic}	%
\usepackage{algorithm}
\usepackage{bm}
\usepackage[caption=false]{subfig}
\usepackage{mathtools}
\usepackage{fullpage}

\newcolumntype{L}[1]{>{\raggedright\arraybackslash}p{#1}}
\newcolumntype{C}[1]{>{\centering\arraybackslash}p{#1}}
\newcolumntype{R}[1]{>{\raggedleft\arraybackslash}p{#1}}

\theoremstyle{plain} %

\def\E{\mathbb{E}}

\def\cbm{{\bm{c}}}

\def\xbm{{\bm{x}}}
\def\zbm{{\bm{z}}}

\def\zbm{{\bm{z}}}

\def\nbm{{\bm{n}}}

\def\Dbm{{\bm{D}}}

\def\Ibm{{\bm{I}}}

\def\thetabm{{\bm{\theta }}}

\def\Dbm{{\bm{D}}}

\def\Ibm{{\bm{I}}}

\def\Ncal{{\mathcal{N}}}

\def\argmin{\mathop{\mathsf{arg\,min}}} %

\newcommand{\norm}[1]{\left\lVert#1\right\rVert}

\begin{document}

\title{Pseudo-MRI-Guided PET Image Reconstruction Method Based on a Diffusion Probabilistic Model}

\author{Weijie Gan, Huidong Xie, Carl von Gall, G\"unther Platsch, Michael T. Jurkiewicz, \\ Andrea Andrade, Udunna C. Anazodo, Ulugbek S. Kamilov, Hongyu An, and Jorge Cabello
\thanks{This work was supported in part by the Academic Medical Organization of Southwestern Ontario (AMOSO; Grant ID: INN20-023) and Physician's Services Incorporated (PSI) Foundation - LRI7767306.}
\thanks{This work involved human subjects in its research. Approval of all ethical and experimental procedures and protocols was granted by the Western University Research Ethics Board (HSREB 104900).}
\thanks{Weijie Gan is with the Department of Computer Science \& Engineering, Washington University in St. Louis, St. Louis, MO, USA.}
\thanks{Huidong Xie is Department of Radiology and Biomedical Imaging, Yale University, New Haven, CT, USA.}
\thanks{Michael T. Jurkiewicz is with Department of Medical Imaging and Department of Medical Biophysics, Western University, 1151 Richmond St, London, ON N6A 5C1, Canada.}
\thanks{Andrea Andrade is with Department of Pediatrics and Department of Clinical Neurological Sciences, Western University, 1151 Richmond St, London, ON N6A 5C1, Canada.}
\thanks{Udunna C. Anazodo is with Lawson Health Research Institute, 268 Grosvenor St, London, ON N6A 4V2, Canada and Neurology and Neurosurgery, Montreal Neurological Institute, McGill University, Montreal, QC, H3A 2B4 Canada.}
\thanks{G\"unther Platsch is with Siemens Healthineers AG, Erlangen, Germany.}
\thanks{Ulugbek S. Kamilov is with the Department of Computer Science \& Engineering and Department of Electrical \& System Engineering, Washington University in St. Louis, St. Louis, MO, USA.}
\thanks{Hongyu An is with the Mallinckrodt Institute of Radiology, the Departments of Neurology, Biomedical Engineering, Electrical \& System Engineering, and with the Division of Biology and Biomedical Sciences, Washington University in St. Louis, MO, USA}
\thanks{Jorge Cabello and Carl von Gall are with Siemens Medical Solutions USA, Inc., 810 Innovation Dr, Knoxville, TN 37932, USA.}
}
\renewcommand\footnotemark{}

\date{}

\maketitle

\begin{abstract}
Anatomically guided PET reconstruction using MRI information has been shown to have the potential to improve PET image quality. However, these improvements are limited to PET scans with paired MRI information. In this work we employed a diffusion probabilistic model (DPM) to infer T1-weighted-MRI (deep-MRI) images from FDG-PET brain images. We then use the DPM-generated T1w-MRI to guide the PET reconstruction. The model was trained with brain FDG scans, and tested in datasets containing multiple levels of counts. Deep-MRI images appeared somewhat degraded than the acquired MRI images. Regarding PET image quality, volume of interest analysis in different brain regions showed that both PET reconstructed images using the acquired and the deep-MRI images improved image quality compared to OSEM. Same conclusions were found analysing the decimated datasets. A subjective evaluation performed by two physicians confirmed that OSEM scored consistently worse than the MRI-guided PET images and no significant differences were observed between the MRI-guided PET images. This proof of concept shows that it is possible to infer DPM-based MRI imagery to guide the PET reconstruction, enabling the possibility of changing reconstruction parameters such as the strength of the prior on anatomically guided PET reconstruction in the absence of MRI.
\end{abstract}

\section{Introduction}

The combination of anatomical information with positron emission tomography (PET) measured data has been long explored as a way to regularize the PET inverse reconstruction problem \cite{ref1}. Magnetic resonance imaging (MRI) can provide exquisite anatomical high detail between different soft tissues and low noise, especially in brain imaging where 3D MRI scans have isotropic spatial resolution. Therefore, MRI has the potential to improve PET image quality by reducing noise, improving spatial resolution and quantification. It also has the potential of reducing the PET scan time or injected activity while retaining the PET image quality. However, access to PET/MRI technology is not as extended as to PET/CT, hence the application of MRI-guided PET (MRIg-PET) reconstruction algorithms is strongly limited.

Anatomically-guided PET reconstruction relies on the premise that there exists a correlation between the anatomical information and functional information. Brain gray matter consists of neurons and dendrites connecting surrounding neurons, while white matter consists of axons (surrounded by fatty myelin) connecting neurons from different regions of the brain. Due to their composition there is 2.5-4.1 higher glucose consumption in the grey matter than in the white matter \cite{ref2}. On the other hand, T1-weighted MRI shows high intensity in fatty tissues like the white matter, and darker intensity in the gray matter. This direct correlation between tissue intensity in MRI and glucose consumption makes FDG brain PET/MRI the usual case where anatomically-guided PET reconstruction methods are tested.

PET image reconstruction using prior information is often formulated using a \textit{maximum-a-posteriori} approach \cite{Mehr17,Schr18}, where the data and prior information are combined in the cost function using a hyperparameter to control the weight of the prior information over the acquired PET data. Even though there have been multiple efforts to develop algorithms to adaptively calculate the hyperparameter based on the measured data, it still often requires manual tuning \cite{Tsai20}.

Inference of MRI from PET imagery is not novel. The estimation of MRI from PET, using florbetapir for amyloid imaging, was explored using generative adversarial networks for quantitative PET analysis \cite{ref3}. The resulting volumetric PET analysis using the generated MRI images compared to the acquired MRI images was similar, even though the generated MRI images were blurry, the soft tissue contrast was low and the overall image quality was poor and unrealistic. Inversely, deep generated FDG-PET images have been synthesized using multiple convolution U-net generators from MRI using the Alzheimer’s Disease Neuroimaging Initiative (ADNI) dataset \cite{ref4}. 

The use of artificial intelligence (AI) models for anatomically guided PET reconstruction has already been employed to create a model that takes as input an OSEM-PET reconstructed image and a paired T1w-MRI image, to produce an anatomically guided PET reconstructed image \cite{ref5}. A downside of this approach is that a model is created using as input reconstructed PET images generated with certain reconstruction parameters (i.e. with or without point spread function [PSF] modeling), hence, if different reconstruction parameters are employed the output of the network is uncertain. In addition, this approach uses a specific MRI sequence, T1w-MRI in the case of \cite{ref5}, preventing its application to other MRI sequences which may be more relevant for other disorders.

Diffusion probabilistic models (DPM) refer to a new class of deep generative models which have recently gained popularity due to the excellent performance (see recent reviews in~\cite{ref6,ref7,ref8}). The key concept behind DPM is to iteratively add noise to an original data point until it becomes a pure noise sample, and then reverse this process by denoising, thereby generating new samples. During this process, DPM learn the data distribution, enabling the synthesis of unseen, high-quality data samples. This approach has demonstrated remarkable results, particularly in generating high-resolution natural images.

Recent studies have shown the potential of DPM in synthesis of PET or MR images~\cite{ref9-mri-diffusion,ref10-mri-diffusion,ref11-pet-diffusion-from-mri,ref12-pet-diffusion-from-mri,ref13-pet-diffusion-from-low-dose,ref14-pet-diffusion-from-low-dose}. For example, DPM has been applied to generate pseudo PET images from the low-dose variants~\cite{ref13-pet-diffusion-from-low-dose,ref14-pet-diffusion-from-low-dose} or corresponding MR images~\cite{ref11-pet-diffusion-from-mri,ref12-pet-diffusion-from-mri}. 

In this work we trained a DPM model with 25 datasets of paired T1w MR images and FDG-PET images reconstructed using a standard iterative parameter (ordered subset expected maximum, OSEM), to generate a pseudo-T1w-MRI (deep-MRI) image that is employed to guide the PET reconstruction (deep-PET) using a \textit{maximum-a-posteriori} (MAP) reconstruction algorithm. The model was then tested in eight datasets from the same cohort not included in the training. This approach allows the parameters of the reconstruction, including the strength of the influence of the anatomical information, to be modified according to the needs of the application, as opposed to prior attempts \cite{ref5}, such as modifying the hyper-parameter or the number of iterations. A detailed description of the study cohort, PET acquisition and reconstruction parameters are summarized in \cite{cabello}.

Furthermore, the model generated with the full dose PET reconstructed images, was evaluated providing as input reconstructed PET images obtained from decimated datasets at 25\% and 5\% of the original datasets. Resulting deep-MRI images were compared with the real MRI images, and more importantly, the resulting deep-PET images were compared to the PET images obtained using the real MRI images. The comparison was performed objectively using a volume of interest analysis, and validated clinically based on observations from experienced physicians. 

The methodology presented in this work can be applied to PET/CT scans and emerging dedicated brain PET scanners, in absence of a paired MRI scan, or also in PET/MRI scans where the MRI contains artifacts, such as patient motion, artifacts coming from coils failures or patients with metallic implants.

\section{Methods}

\begin{figure*}
    \centering
    \includegraphics[width=\textwidth]{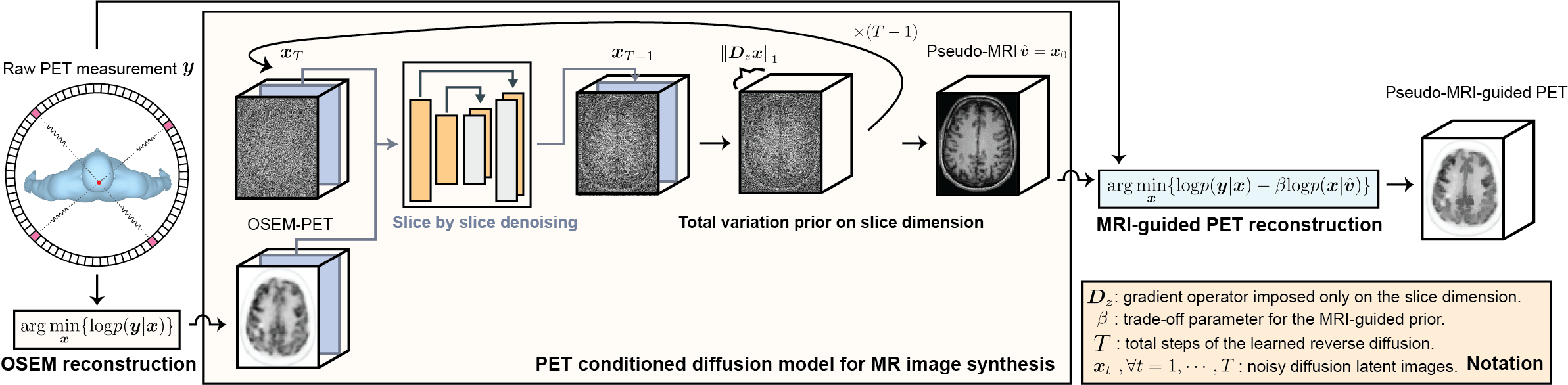}
    \caption{An illustration of the proposed deep-MRI-guided PET image reconstruction. Firstly, we generate a deep-MRI image from the OSEM PET image by using a pre-trained diffusion model (see its training process in Section~\ref{sec:DPM} and Section~\ref{sec:cDPM}). Then, the deep-MRI-guided PET image is obtained by solving an optimization of the anatomically-guided reconstruction given the generated deep-MRI images and the raw PET data.}
    \label{fig:pipeline}
\end{figure*}

\subsection{Diffusion Probabilistic Model}
\label{sec:DPM}

DPM consists of two Markov processes: the fixed forward process and the learning-based reverse process. The forward process starts from a sample of a T1w-MRI image $\xbm_0\sim q(\xbm)$ and gradually adds Gaussian noise according to the following transition probability:
\begin{equation}
    q(\xbm_t|\xbm_{t-1}) := \Ncal(\xbm_t;\sqrt{1-\beta_t}\xbm_{t-1},\beta_t\textbf{I})\ ,
\end{equation}
where $\Ncal(\cdot)$ represents the Gaussian probability density function, and $\beta_{1:T}$ refers to a variance schedule subject to $\beta_t\in(0,1)$ for all $t=1,...,T$. An important feature of this forward process is that it allows us to sample at any arbitrary step $t$ directly conditioned on $\xbm_0$
\begin{equation}
    \label{q:x_t}
    q(\xbm_t|\xbm_0) := \Ncal(\xbm_t;\sqrt{\bar{\alpha_t}}\xbm_0, (1-\bar{\alpha_t})\textbf{I})\ ,
\end{equation}
where $\alpha_t := 1-\beta_t$ and $\bar{\alpha}_t=\prod_{s=1}^t\alpha_s$.
The latent variables $\xbm_{1:T}$ have the same dimensionality as the T1w-MRI image. The latent $\xbm_T$ is nearly an isotropic Gaussian distribution for a properly designed $\beta_t$ schedule. Since the latent $\xbm_T$ is a known isotropic Gaussian distribution, one can generate a new $\xbm_T$ and then obtain a T1w-MRI image $\xbm_0$ by progressively sampling from the reverse posterior $q(\xbm_{t-1}|\xbm_t)$.
However, this reverse posterior is tractable only if $\xbm_0$ is known
\begin{equation}
\label{equ:q-posi}
    q(\xbm_{t-1}|\xbm_t, \xbm_0) = \Ncal\Big(\xbm_{t-1}; \mu_q(\xbm_t,\xbm_0), \frac{\beta_t(1-\bar{\alpha}_{t-1})}{1-\bar{\alpha}_t}\textbf{I}\Big)\ ,
\end{equation}
where
\begin{equation}
    \mu_q(\xbm_t,\xbm_0) = \frac{\sqrt{\alpha_t}(1-\bar{\alpha}_{t-1})\xbm_t + \sqrt{\bar{\alpha}_{t-1}}(1-\alpha_t)\xbm_0}{1-\bar{\alpha}_t}\ .
\end{equation}
Noted that $q(\xbm_{t-1}|\xbm_t):=q(\xbm_{t-1}|\xbm_t, \xbm_0)$, where the extra conditioning term $\xbm_0$ is superfluous due to the Markov property.
DPM proposes to learn a parameterized Gaussian transitions $p_\thetabm(\xbm_{t-1}|\xbm_t)$ to approximate \eqref{equ:q-posi}
\begin{equation}
\label{equ:p-posi}
    p_\thetabm(\xbm_{t-1}|\xbm_t) = \Ncal\Big(\xbm_{t-1}; \mu_\thetabm(\xbm_t,t), \sigma^2_t\textbf{I}\Big)\ ,
\end{equation}
where
\begin{equation}
    \label{equ:p-mean}
    \mu_\thetabm(\xbm_t,t) = \frac{1}{\sqrt{\alpha_t}}\Big(\xbm_t-\frac{1-\alpha_t}{\sqrt{1-\bar{\alpha}_t}}\epsilon_\thetabm(\xbm_t, t)\Big)\ .
\end{equation}
Here, $\epsilon_\thetabm(\xbm_t, t)$ denotes a neural network. The training objective of $\epsilon_\thetabm(\xbm_t, t)$ can be formulated as follow by considering a training set of $\{\xbm_i\}_i$
\begin{equation}
    \label{equ:training}
    \E_{\xbm,\epsilon,t\sim[1,T]}\big[\norm{\epsilon - \epsilon_\thetabm(\xbm_t, t)}^2\big]\ ,
\end{equation}
where $\xbm_t$ denotes a noisy latent sampled from eq. \eqref{q:x_t} as $\xbm_t=\sqrt{\bar{\alpha_t}}\xbm+\sqrt{(1-\bar{\alpha_t})}\epsilon$ for $\epsilon\sim\Ncal(0,\textbf{I})$.
Recent studies~\cite{dhariwal2021diffusion,nichol2021improved} have shown the improved performance by using the learned variance $\sigma_t^2\coloneqq\sigma_\thetabm^2(\xbm_t, t)$. We also adopted this approach. To be specific, we have $\sigma_\thetabm(\xbm_t, t)\coloneqq\exp(v\log \beta_t + (1-v)\log\tilde{\beta}_t)$, where $\tilde{\beta}_t$ refers to the lower bounds for the reverse diffusion posterior variances \cite{ho2020denoising}, and $v$ denotes the network output. We used a single neural network with two separate output channels to estimate the mean and the variance of \eqref{equ:p-posi} jointly.
Based on the learned reverse posterior $p_\thetabm(\xbm_{t-1}|\xbm_t)$, the iteration of obtaining a $\xbm_0$ from a $\xbm_T$ can be formulated as 
\begin{equation}
    \xbm_{t-1} = \mu_\thetabm(\xbm_t,t) + \sigma_t\zbm,\text{ where } \zbm\sim\Ncal(0,\textbf{I})\ .
\end{equation}

\subsection{PET-conditioned MRI Synthesis Diffusion Model}
\label{sec:cDPM}

The conventional diffusion models discussed in Section~\ref{sec:DPM} are for unconditional MR image synthesis, namely generating arbitrary realistic MR image. In this study, we reconstructed a set of PET images using the OSEM algorithm, and then trained a diffusion model using paired OSEM-PET reconstructed images and T1w-MRI images, in order to generate pseudo MR images that are consistent with the OSEM-PET images. To be specific, let $\cbm$ denote the OSEM-PET images. Our goal is to learn a new parameterized Gaussian transitions $p_\thetabm(\xbm_{t-1}|\xbm_t,\cbm)$ to approximate eq. \eqref{equ:q-posi} instead of eq. \eqref{equ:p-posi} in the original DPM. This approach can be achieved by simply supplying $\cbm$ as the additional network input. The new training objective can then be formulated as follows by considering a training set MRI-PET image pairs $\{(\xbm_i,\cbm_i)\}_i$
\begin{equation}
    \E_{(\xbm,\cbm),\epsilon,t\sim[1,T]}\big[\norm{\epsilon - \epsilon_\thetabm(\xbm_t, t, \cbm)}^2\big]\ .
\end{equation}
Based on $p_\thetabm(\xbm_{t-1}|\xbm_t,\cbm)$, the iteration of the PET-conditioned DPM becomes 
\begin{equation}
    \xbm_{t-1} = \mu_\thetabm(\xbm_t,t,\cbm) + \sigma_t\zbm,\text{ where } \zbm\sim\Ncal(0,\textbf{I})\ ,
\end{equation}
where
\begin{equation}
    \mu_\thetabm(\xbm_t,t, \cbm) = \frac{1}{\sqrt{\alpha_t}}\Big(\xbm_t-\frac{1-\alpha_t}{\sqrt{1-\bar{\alpha}_t}}\epsilon_\thetabm(\xbm_t, t, \cbm)\Big)\ .
\end{equation}

\subsection{3D Image Synthesis using 2D Diffusion Model}
PET and MRI are 3D imaging modalities, yet directly synthesizing an entire 3D MR volume with a diffusion model poses practical challenges due to the extremely high memory and computational cost. An alternative approach involves training a diffusion model for 2D PET-conditioned MR image synthesis and subsequently generating the entire 3D MR volume in a \emph{slice-by-slice} manner.
Although effective, this simple approach may lead to shaggy imaging artifacts along the slice dimension~\cite{pretrained2d-tv, pretrained2d-huidong, pretrained2d-hyy}. This issue arises because direct slice-by-slice synthesis fails to consider the spatial correlation of the 3D volume along the slice dimension. To mitigate such artifacts, we adapted~\cite{pretrained2d-tv} to incorporate an optimization strategy into each diffusion sampling step, leveraging a total variation (TV) prior exclusively imposed along the slice direction. This enforces spatial consistency in the 3D volume, particularly in the slice direction.
To specify, let ${\bm C}(x,y,z)$ represent a 3D PET volume, and ${\bm X_T}(x,y,z)$ denote a 3D latent variable sampled from an isotropic Gaussian distribution, where $x$, $y$, and $z$ denote physical coordinates. The diffusion sampling process iteratively updates the entire ${\bm X_t}(x,y,z)$ volume for $t$ ranging from $T$ to $0$.
In each iteration, each slice of ${\bm X_t}(x,y,z)$ is independently updated as follows
\begin{equation}
{\bm Y_{t-1}}(x,y,z_i) = \mu_\thetabm({\bm X_{t}}(x,y,z_i),t,{\bm C}(x,y,z_i))+\sigma_t\nbm,
\end{equation}
where $\nbm\sim\Ncal(0, \Ibm)$, and $z_i=1,\cdots,N_z$ with $N_z$ being the total number of slices. Note that ${\bm X_{t}}(x,y,z_i)$ and ${\bm C}(x,y,z_i)$ are 2D images, allowing the use of standard 2D diffusion models.
Subsequently, we solve the optimization problem below to obtain ${\bm X_{t-1}}(x,y,z)$
\begin{equation}
\label{equ:tv}
\begin{aligned}
& {\bm X_{t-1}}(x,y,z) = \\ & \argmin_{{\bm X}(x,y,z)} \norm{{\bm X}(x,y,z)-{\bm Y_{t-1}}(x,y,z)}_2^2 + \tau\norm{\Dbm_z {\bm X}(x,y,z)}_1
\end{aligned}
\end{equation}
where $\Dbm_z$ denotes a gradient operator along the z dimension, and $\tau$ is a trade-off parameter. In practice, we use gradient descent to solve this optimization problem and optimize $\tau$ through grid-search on a validation set.

\subsection{Anatomically-guided Reconstruction Method}

The MRIg-PET image reconstruction method used in this work was the Bayesian \textit{maximum-a-posteriori}, including the MRI information modeled as a Markov random field. The prior model was formulated as 

\begin{equation}
 R(u)=\sum^N_j\phi ( \sum_{b\in N_j} \xi_{jb}\omega_{jb}\psi(u_j - u_b) ),
\end{equation}

where $\phi$ is the potential function operating in $N_j$ around voxel index $j$, $\xi_{jb}$ is a coefficient defined as $1/2N_j\sigma^2_u$ ($\sigma_u$ is the standard deviation in the neighborhood $N_j$ in the PET image $u$), $\omega_{jb}$ is a similarity coefficient, and $\psi$ is the quadratic similarity relation. $\psi$ is weighted by $\omega_{jb}$ based on the Burg joint entropy between the PET and MRI images \cite{Mehr17}, defined as 

\begin{equation}
 \omega_{jb}=\frac{1}{p(u_j,v_j)}~exp(-\frac{(u_j - u_b)^2}{2\sigma_u^2})~exp(-\frac{(v_j - v_b)^2}{2\sigma^2_{v}}).
 \label{eq2}
\end{equation}

where \textbf{\textit{v}} is the MRI image, $\sigma$ is the standard deviation in $N_j$ in \textbf{\textit{u}} and \textbf{\textit{v}} , and $p(u_j,v_j)$ is defined as a non-parametric Parzen window. The weights are constrained to only those voxels which fulfill a certain level of similarity \cite{bowsher}. The optimization was written as a preconditioned gradient ascend \cite{nuyts}. 

The derivative of $R(u)$ is included in the reconstruction with a hyper-parameter which balances the influence of the prior and the measured data. The hyper-parameter was individually recalculated at each iteration ($k$) for each voxel ($j$), depending on the image noise following

\begin{equation}
 \beta^k_j=\alpha \sqrt{z^{k-1}_j \sum_{i=1}^M g_{ij}n_ia_i},
\end{equation}

where $\alpha$ is a constant (1 in this work), $g$ is the system model, $n$ is the normalization and $a$ is the attenuation. The weight $z^{k-1}_j$ is calculated based on the MRI and PET information as follows: First, the MRI image intensity is scaled to the dynamic range of the PET image at each iteration. Secondly, each voxel is filtered by calculating the mean intensity inside the region of interest determined by similar intensities in the anatomical image \cite{cabello}.

\subsection{Effect of Decimated Datasets}

One of the potential applications of anatomically-guided PET reconstruction is the possibility of reducing the scan time or injected activity in the subject. Therefore, low counts applications render to investigate the performance of the AI model with reduced counts datasets. For that purpose, we randomly decimated the acquired listmode files, from 100\% of the counts, down to 25\% and 5\% of the original counts in the eight subjects used for validation. The validation was performed visually, as well as quantitatively via volume of interest (VOI) analysis using Statistics Parametric Mapping (SPM12) \cite{spm}. The analysed VOIs comprise a range of cortical and sub-cortical regions of different sizes.

\subsection{Subjective Image Quality Evaluation}

Reconstructed PET images using OSEM and MRIg-PET (with the acquired and the deep-MRI images) of the eight datasets used for evaluation were inspected by two board-certified nuclear medicine physicians with $>$10 years of experience. All the images were evaluated using a 4-point scale (excellent, good, poor, nondiagnostic) on noise, sharpness, general image quality, and confidence. The evaluation was performed on the three dose levels: 100\%, 25\% and 5\%. A paired t-test was performed between patients scored by both physicians reconstructed with the three methods to determine the statistical significance.

\section{Results}
\subsection{MRI Analysis}

\begin{figure}[ht]
\centering
\includegraphics[scale=0.35]{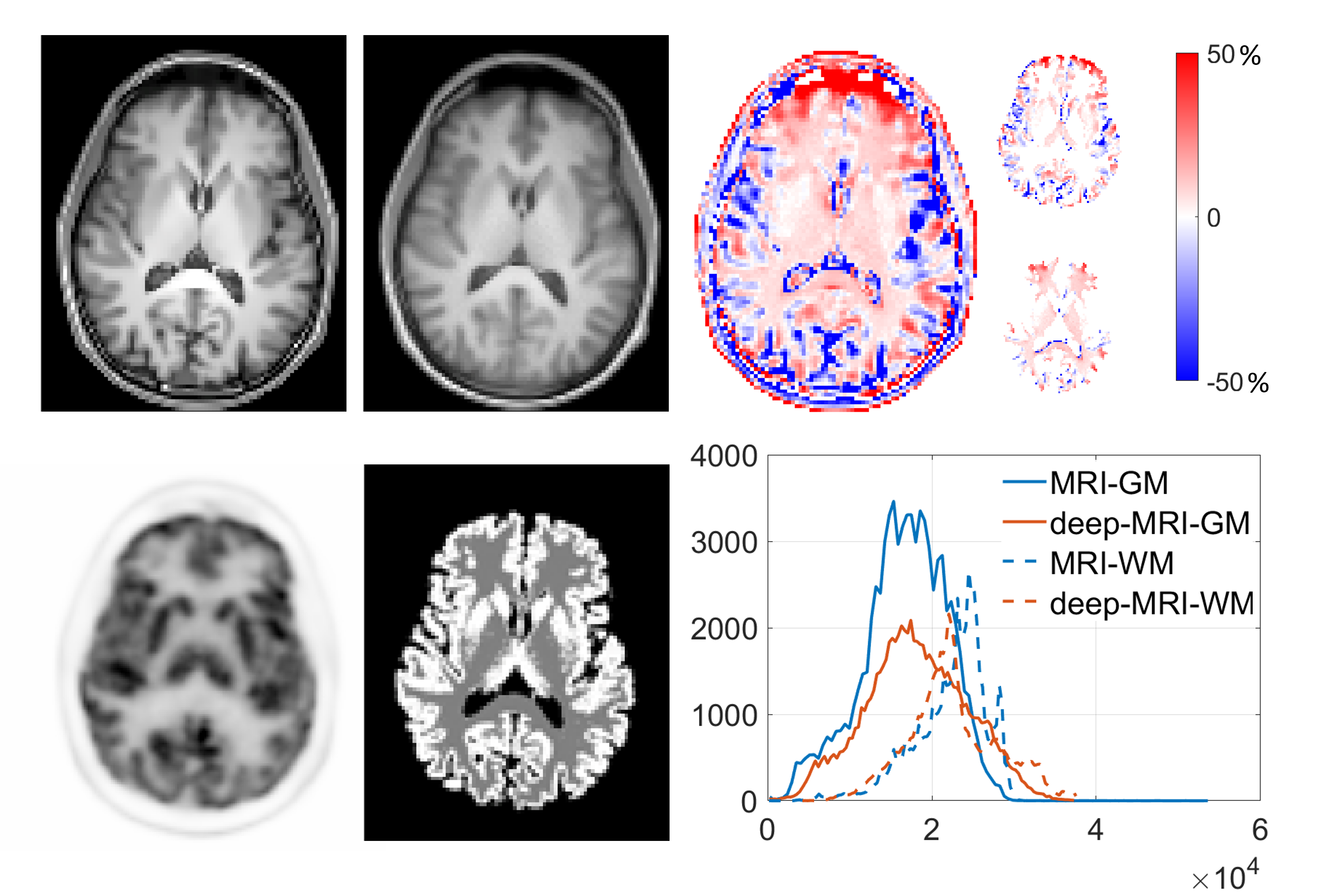}
\caption{MRI acquired from scanner (top left), deep-MRI (top center), relative difference (top right), reconstructed PET image using OSEM (bottom left), gray and white matter masks (bottom center) and histogram of gray and white matter from acquired and deep-MRI images (bottom right) of an exemplar subject.}
\label{fig:mri1}
\end{figure}

Figure \ref{fig:mri1} shows an example of an acquired T1w-MRI image and its paired result from the DPM, together with the PET reconstructed image using OSEM as input to the model, the relative difference between both MRI images, and the relative difference masked by the gray and white matter masks. The masks of the segmented gray and white matter of the acquired T1w-MRI is shown in figure \ref{fig:mri1}, together with the histograms of the gray and white matter from the acquired and deep-MRI images.

Visually, both MRI images show high structural resemblance, but the acquired MRI shows better soft tissue contrast and spatial resolution than the deep-MRI. The relative difference shows that the highest error is visible outside the brain area, where voxels with low intensity values are predominant. Overall, the mean relative differences in the gray and white matter between the acquired MRI and the deep-MRI are -3.4\% and 1.4\% respectively. The histograms show similar shapes but the histogram corresponding to the deep-MRI image confirms the aforementioned reduced contrast between gray and white matter.

\subsection{Qualitative PET Image Quality Comparison}

\begin{figure}[ht]
\centering
\includegraphics[scale=0.35]{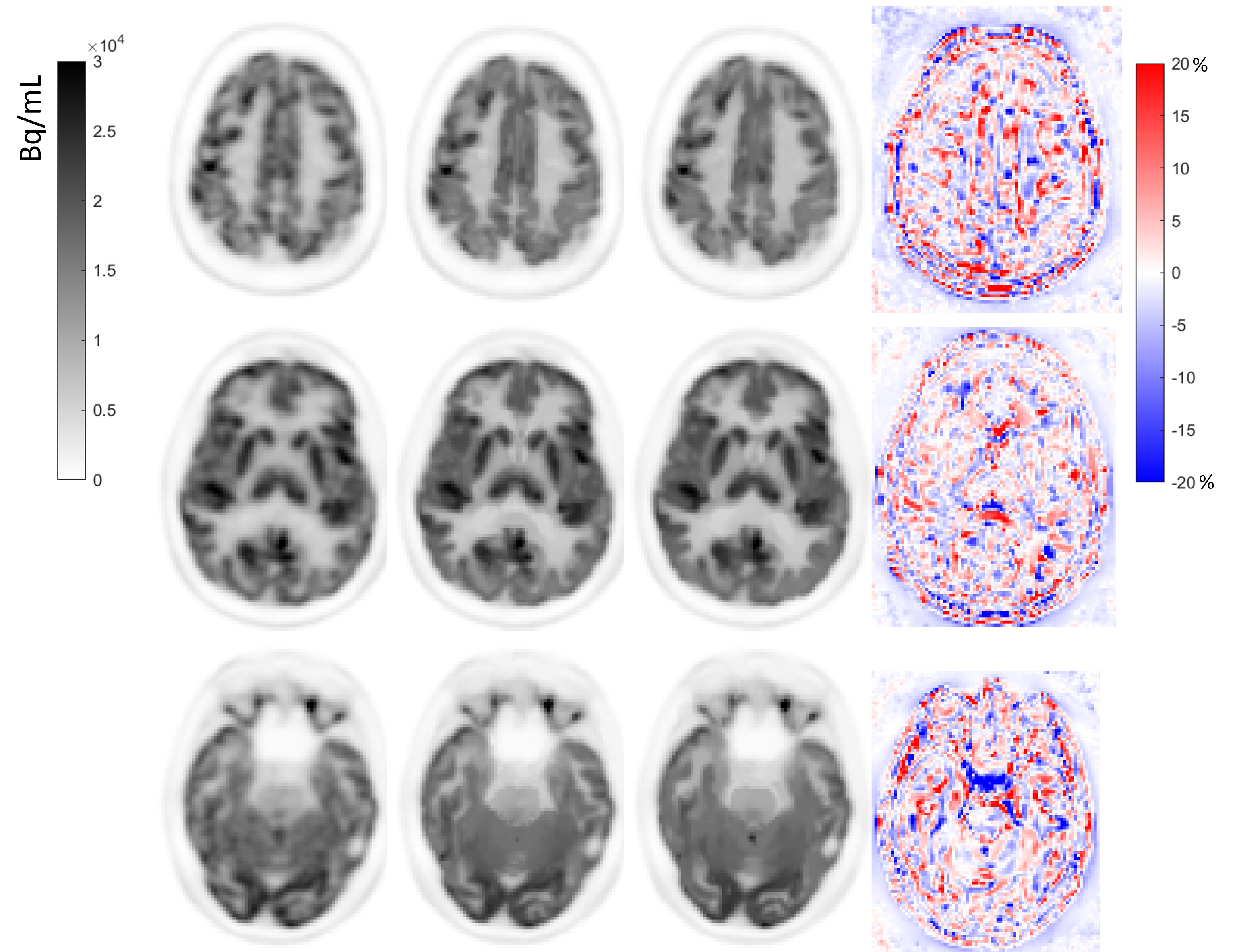}
\caption{Axial slices at different levels of reconstructed PET image using OSEM (first column), MRIq-PET image obtained using the acquired MRI (second column), MRIg-PET image obtained using the deep-MRI (third column), and relative difference between the MRIg-PET images obtained with the acquired and deep-MRI images.}
\label{fig:pet1}
\end{figure}

Figure \ref{fig:pet1} shows the axial slices of the PET reconstructed images obtained with OSEM and MRIg-PET using the acquired and deep-MRI images, together with the relative difference measured between the two PET reconstructed images using the acquired and deep-MRI images. Both MRIg-PET images show high similarities, and both show reduced noise and higher sharpness than the OSEM image. Visually, relative differences do not exceed $\sim\pm$20\%, showing the largest differences in areas with low PET signal.

\subsection{PET Volume of Interest Analysis}
\label{sec:voisec}

\begin{figure}[ht]
\centering
\includegraphics[scale=0.45]{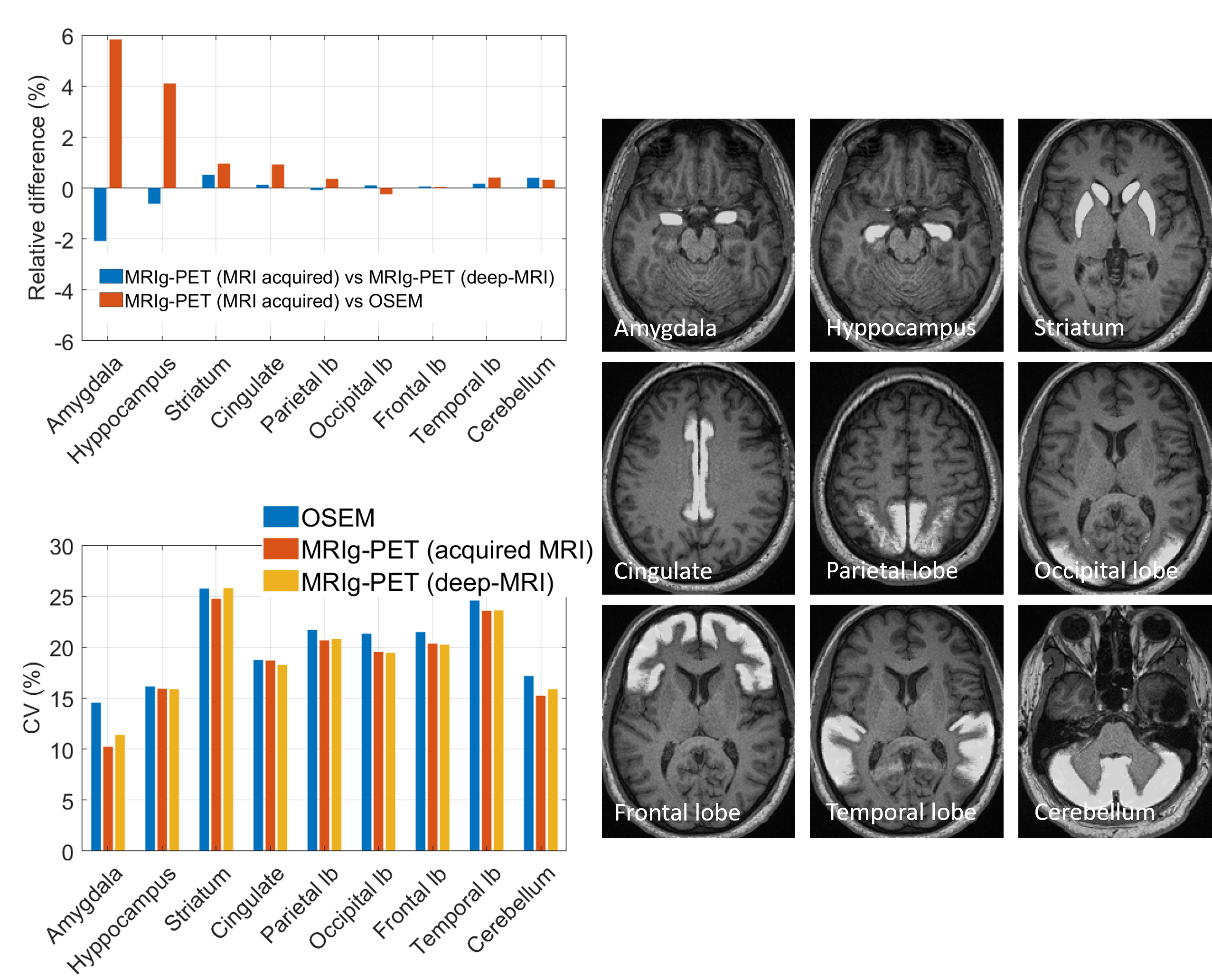}
\caption{Mean relative difference between MRIg-PET reconstruction using the acquired and deep-MRI images and between MRIg-PET reconstruction using the the acquired MRI and OSEM for each ROI, averaged between eight subjects (top left). CV measured in each ROI obtained with MRIg-PET reconstruction using the acquired and deep-MRI images, and OSEM, averaged between eight subjects (bottom left). }
\label{fig:voianl}
\end{figure}

We measured the mean and standard deviation in each VOI of the eight subjects used for validation, normalized to the Hammers atlas in MNI space. Figure \ref{fig:voianl} shows the mean relative difference between the mean values measured in each ROI between the MRIg-PET images using the acquired and the deep-MRI images, and between the MRIg-PET image and OSEM images. Figure \ref{fig:voianl} also shows the mean coefficient of variation (CV), measured in each VOI for OSEM and MRIg-PET using the acquired and the deep-MRI images. The VOIs used in the analysis are shown on the right side of figure \ref{fig:voianl}.

The VOIs in figure \ref{fig:voianl} are presented in order of increasing volume from the amygdala ($\sim$2.3 mL) up to the cerebellum ($\sim$114 mL). The relative differences show a correlation with volume size, showing large differences between mean intensities obtained with the different reconstructions for the smallest VOIs, and small differences for the larger regions. The differences between MRIg-PET using the different MRI images show smaller differences than between MRIg-PET images with the acquired MRI and OSEM. The CVs obtained with MRIg-PET, independently of the MRI image used, were consistently lower than with OSEM.

\subsection{Performance with Decimated Datasets}
\label{sec:decim}

\begin{figure}[ht]
\centering
\includegraphics[scale=0.35]{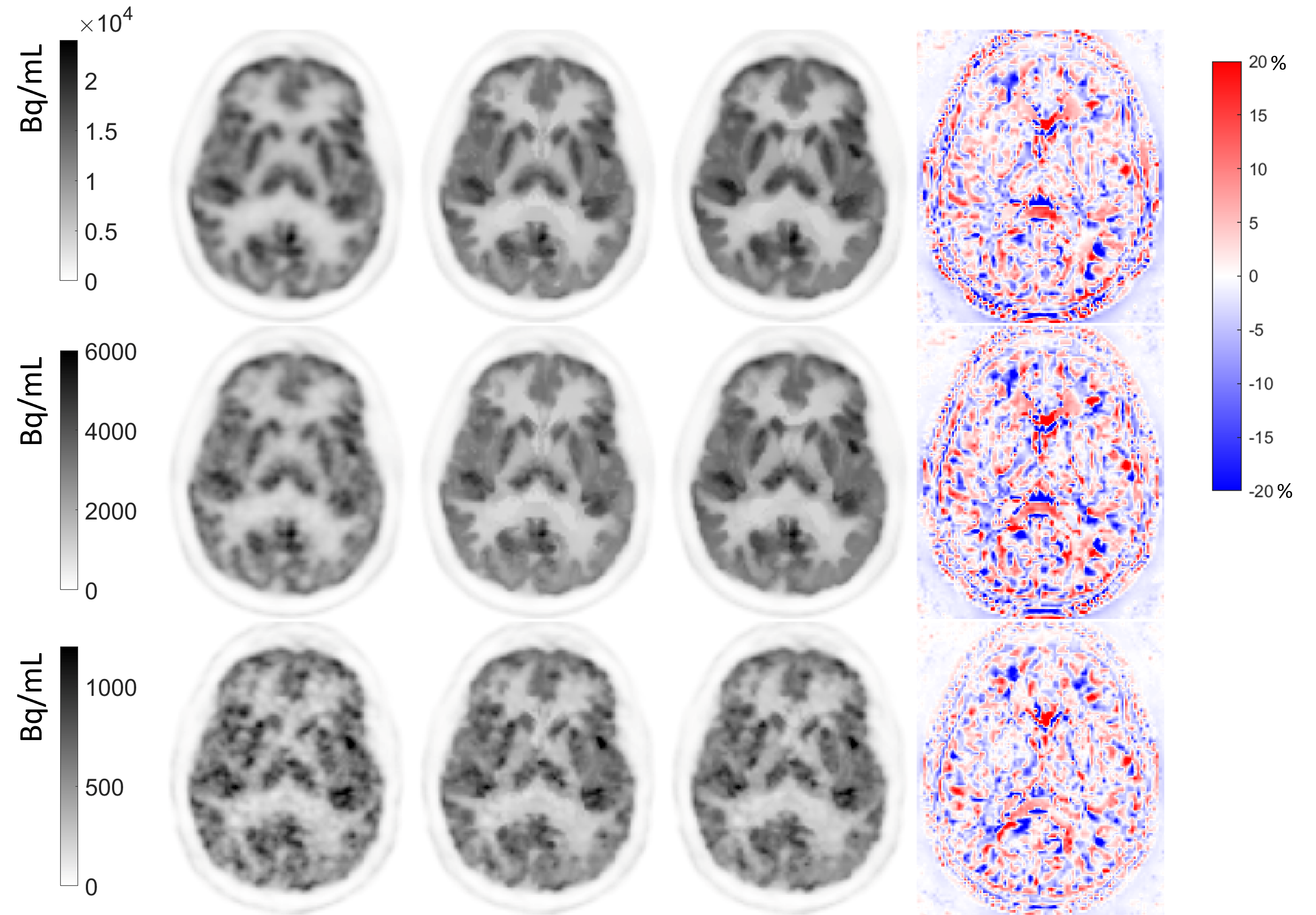}
\caption{Axial slices of PET reconstructed images from the full dataset (first row), 25\% of the counts (second row), 5\% of the counts (third row), reconstructed with OSEM (left column), and with MRIg-PET using the acquired MRI (second column) and the deep-MRI (third column). The relative difference between the PET reconstructed images using MRIg-PET with the measured and deep-MRI are shown in the fourth column. }
\label{fig:doses}
\end{figure}

Figure \ref{fig:doses} shows the reconstructed PET axial slices of the same patient as figure \ref{fig:pet1}, using 100\% of the acquired counts, and the corresponding images obtained with 25\% and 5\% of the acquired counts.

Visually the MRIg-PET reconstructed images using the acquired and deep-MRI show high resemblance. The relative differences show errors up to 20\% in the cold regions, mainly where the CSF is located, and lower in the hot and warm regions, corresponding to the gray and white matter. The MRIg-PET reconstructed images show overall similar improvements compared to the OSEM-reconstructed image in terms of noise and image sharpness.

\begin{figure}[ht]
\centering
\includegraphics[scale=0.4]{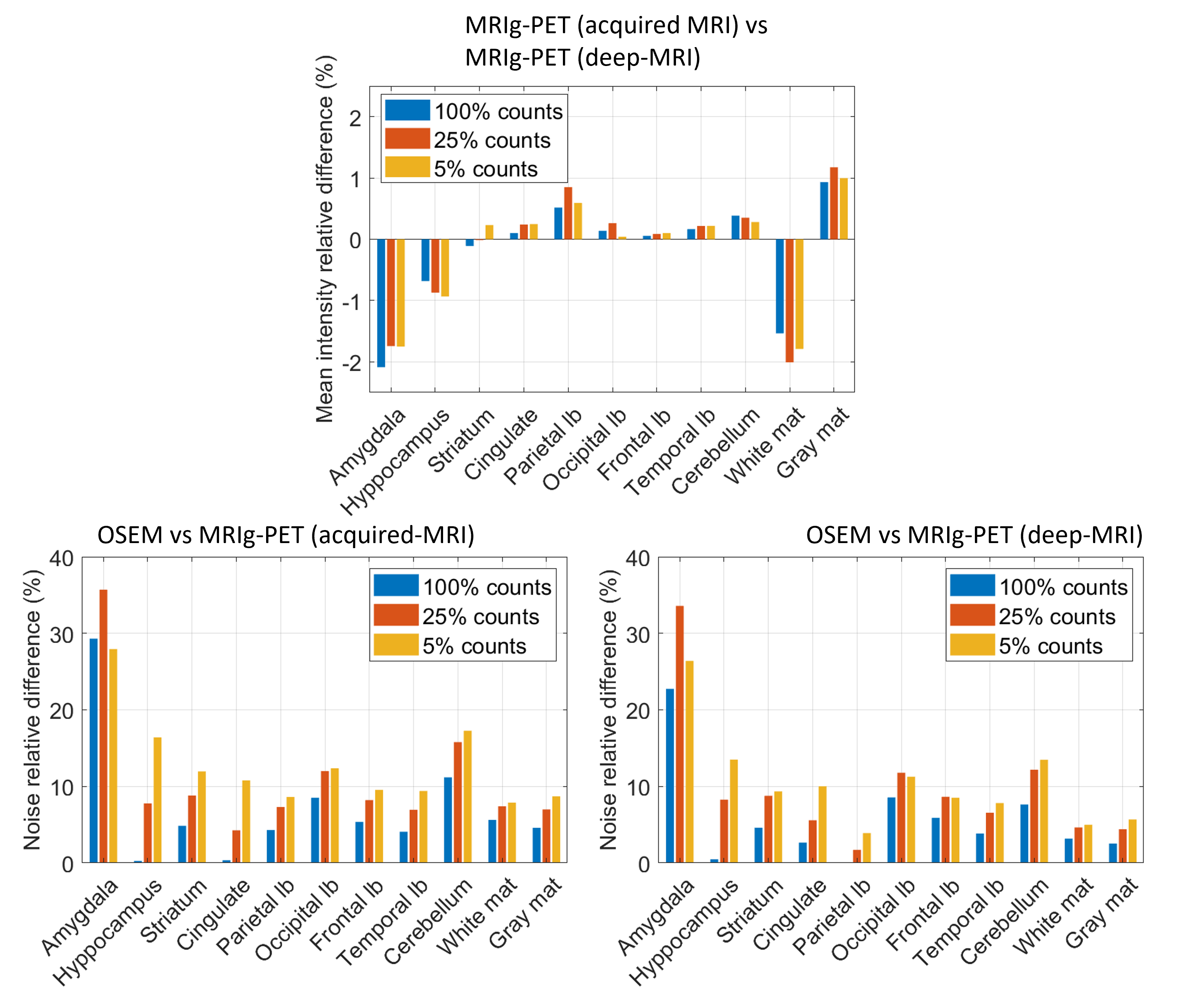}
\caption{Mean intensity difference between PET images obtained with the acquired MRI and deep-MRI for different regions, obtained for 100\%, 25\% and 5\% of the measured counts (top). Noise difference between OSEM and the MRIg-PET reconstructed image obtained with the acquired MRI (bottom left) and deep-MRI (bottom right) for different regions, obtained for 100\%, 25\% and 5\% of the measured counts.}
\label{fig:fig6}
\end{figure}

Figure \ref{fig:fig6} shows the relative difference between mean intensities measured in multiple VOIs, between the PET reconstructed images obtained with the acquired MRI and the deep-MRI. In addition, the improvement in noise of each MRI-guided PET reconstructed image compared to OSEM is also shown in figure \ref{fig:fig6}. Comparing the mean intensity in the VOIs of the PET reconstructed images obtained with the measured and deep-MRI images, the largest differences are correlated with the VOI volume (with the exception of the gray and white matter), showing the smallest regions the largest differences. The largest differences were measured in the amygdala ($\sim$2\%), followed by the gray and white matter. We did not observe systematic differences between the different levels of counts in the datasets, suggesting that the quality of the deep-MRI is relatively independent on the PET image noise.

The noise measured between OSEM and the MRIg-PET images (with both MRIs) was similar for all the VOIs, showing the largest difference in the amygdala, and consistently having higher improvements between OSEM and either MRIg-PET method for the lowest level of counts and lower improvement for the highest level of counts.

\subsection{Subjective Image Quality Evaluation}

\begin{figure}[ht]
\centering
\includegraphics[scale=0.4]{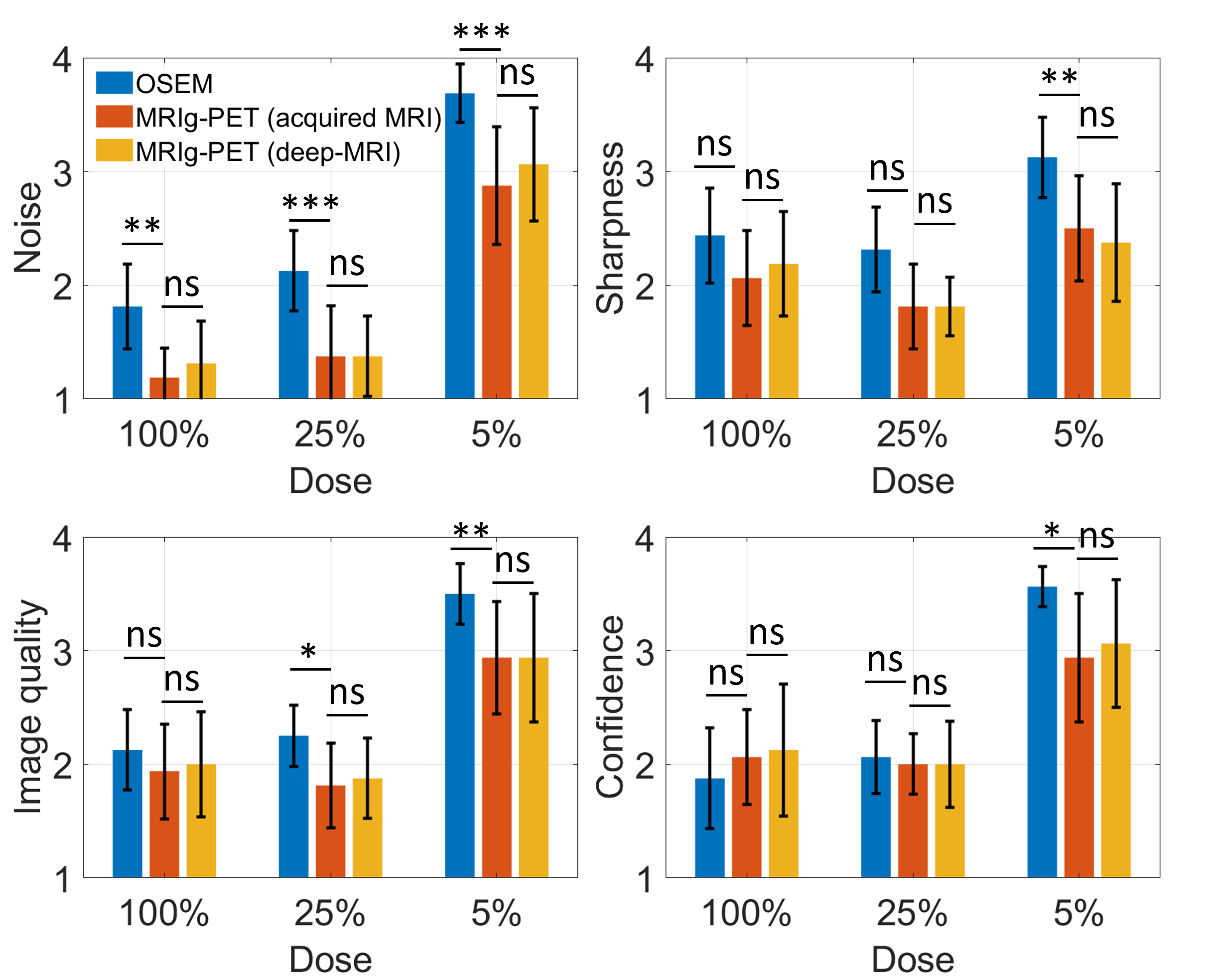}
\caption{4-point scale on noise, sharpness, image quality and confidence evaluated for OSEM, MRIg-PET using the acquired MRI, and MRIg-PET using the deep-MRI, for three dose levels. The statistical significance between OSEM and MRIg-PET with the acquired MRI, and between the two MRIg-PET images is shown for each plot.}
\label{fig:fig7}

\end{figure}

In order to confirm if the quantitative measurements from Sections~\ref{sec:voisec} and \ref{sec:decim} can be transferred to visual improvement from a clinical perspective, the images of the eight subjects used for validation were inspected by two physicians. The results obtained from the two physicians and the eight patients were combined to extract the mean and standard deviation for each reconstruction algorithm and each dose level. Figure \ref{fig:fig7} shows the resulting combined scores for noise, sharpness, image quality and confidence, for the three dose levels: 100\%, 25\% and 5\%. The statistical significance is indicated as ns (no significance), * (p\textless0.05), ** (p\textless0.01) and *** (p\textless0.001).

PET images reconstructed with OSEM were scored as good (2) for all the figures of merit with 100\% and 25\% of the counts, while MRIg-PET images were scored as excellent (1) for noise and as good (2) for sharpness, image quality and confidence, for 100\% and 25\% of the counts. However, MRIg-PET images were overall scored consistently slightly higher (lower score) than OSEM images. MRIg-PET shows overall superior performance compared to OSEM in noise, sharpness and image quality as shown in figure \ref{fig:fig7}. However, the three algorithms show similar confidence for the 100\% and 50\% dose levels. With respect to the two MRIg-PET algorithms, the physicians rated the images similarly, showing in some cases better scores for the MRIg-PET reconstruction using the acquired MRI, but statistically not significant. Most differences were found between OSEM and MRIg-PET in the case of noise for all the dose levels. Additionally, significant differences were found between OSEM and MRIg-PET for all the figures of merit for the lowest dose level of 5\%.
 
\subsection{Results with Defective MRI}

The poor quality in MRI images may happen due to subject motion or hardware problems. For instance, the MRI-T1w sequence takes ~5-min to be acquired. During that time, the subject may move the head causing motion artifacts and resulting in poor image quality. Such images may potentially introduce artifacts in the PET images, when combined during the MRI-guided PET reconstruction process. 

\begin{figure}[ht]
\centering
\includegraphics[scale=0.35]{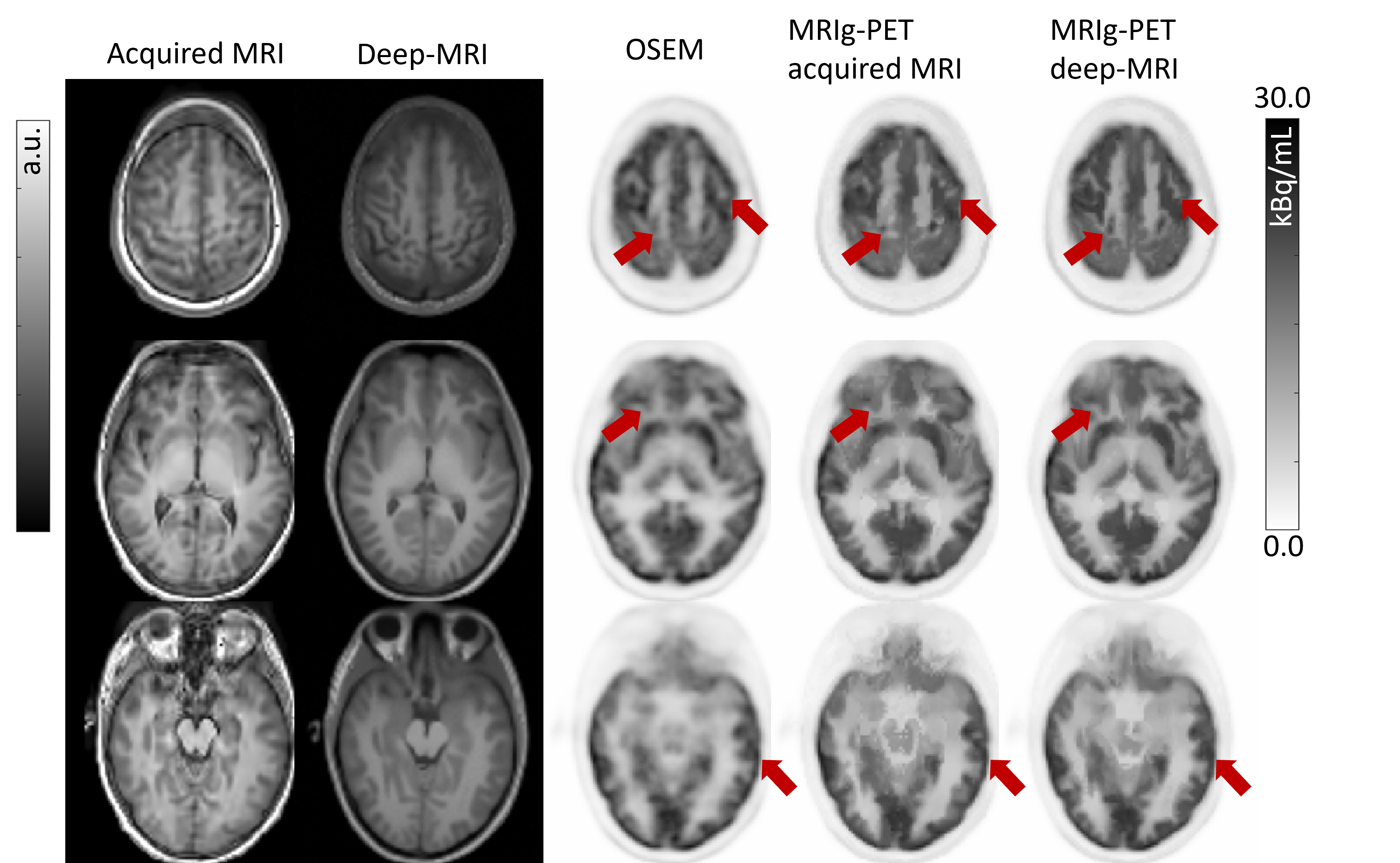}
\caption{Axial slices of an exemplar subject who moved during the acquisition of the MRI-T1w sequence. The first column is the originally acquired MRI-T1w images, the second column is the deep-MRI-T1w images, the third column is the PET reconstructed images using OSEM, the fourth column is the MRIg-PET reconstructed image with the acquired MRI-T1w image, and the fifth column is the MRIg-PET reconstructed image with the deep-MRI-T1w image}
\label{fig:defect}
\end{figure}

Figure \ref{fig:defect} shows an example of a 12-year old male which shows strong artifacts in the MRI-T1w images due to motion, which are transferred to the MRIg-PET image. The deep-MRI image shows lower image contrast but no motion artifacts. Therefore, the MRIg-PET reconstructed image obtained with the deep-MRI image does not show artifacts, and shows higher resemblance to the OSEM images, compared to the MRIg-PET image based on the acquired MRI image. Some regions with visible artifacts in the MRIg-PET images obtained with the acquired MRI image, are indicated with arrows in the reconstructed PET images.

\section{Discussion}

The use of low noise and 3D high spatial resolution MRI to guide the PET reconstruction has extensively been investigated to regularize the ill-posed inversion problem of PET image reconstruction. There are a considerable amount of algorithms that formulate the problem in different ways. However, all those methods require a paired MRI with a high level of correlation between the functional or metabolic information from PET and anatomical information from MRI, which greatly limits the application of MRI-guided PET reconstruction.

Diffusion probabilistic models (DPM) are emerging as a method to produce natural realistic images with high spatial resolution. Therefore, we employed in this work DPM to generate highly realistic MRI images from nothing else than PET images as input, in order to use as reference to guide the PET image reconstruction, rather than for any clinical purpose. The premise is that there needs to be be a high correlation between the patterns in the PET and the MRI images, as is usually the case in FDG-PET and MRI-T1w.

In MRI-guided PET reconstruction, a certain level of information is transferred from the MRI image to the PET image. If too much information is transferred, the PET image may resemble an MRI image while if too little information is transferred, the PET image will be similar to the OSEM-PET image. Usually, the level of transferred information is limited in order to keep the image as PET-like as possible, while improving the image quality.  The purpose of this work was to generate an MRI image of enough realism and quality to be able to guide the PET image reconstruction without introducing hallucinations, not as accurate as the acquired MRI images, but still improving the PET image quality. 

The resulting MRI-T1w images obtained from DPM highly resembled the acquired MRI-T1w images, showing in general similar noise properties. Spatial resolution was slightly degraded and contrast between soft tissues decreased in the deep-MRI images compared to the acquired MRI images. However, by inspecting the MRIg-PET images obtained with the acquired MRI and the deep-MRI, small differences were observed, suggesting that the level of degradation in the MRI images was not enough to be transferred to the PET images, while still improving the PET image quality. This was confirmed by a VOI analysis performed in multiple cortical and subcortical brain regions. 

Similar findings were observed in decimated datasets down to 25\% and 5\% from the original number of counts. The largest differences in mean intensity between MRIg-PET images with the acquired and deep-MRI were observed to be $\sim$2\% in the amygdala and white matter. Comparing the noise measured in all the regions between OSEM and both MRIg-PET, the improvement introduced by MRIg-PET was similar in all the regions.

To confirm if the quantitative observations could be transferred to actual clinical readings, two physicians blindly evaluated all the images using a 4-point score on noise, sharpness, image quality and confidence on the reading. Their observations confirmed that the MRIg-PET images obtained with the acquired and the deep-MRI showed high resemblance in the four figures of merit, with no statistical significance between images. Comparing to OSEM, the only consistent observations were statistically significant improvement of MRIg-PET compared to OSEM in noise, and for the four figures of merit measured on the images reconstructed with 5\% of the original counts. Besides statistical measurements, MRIg-PET images in general scored better (lower score - better image quality) than OSEM images. The more similar scores between OSEM and MRIg-PET were obtained in the reading confidence.

Finally we also highlighted the benefits of using a deep-MRI in PET/MRI cases where the MRI was defective. While hardware problems are not common, patient motion is often encountered, especially in patients who undergo neurological disorders. The MRIg-PET obtained with deep-MRI showed higher resemblance to the OSEM image compared to the acquired MRI in terms of hallucinations, while preserving the expected improved image quality from MRIg-PET compared to OSEM. It is important to highlight that patient motion during the PET measurement was not addressed in this work, and remains future work. It is also important to note that subject motion is a more critical problem to address when using anatomical information in the reconstruction compared to reconstruction algorithms without anatomical guidance, since misregistrations can potentially introduce strong artifacts in the PET reconstructed images.

This study considers a relatively small PET/MRI training dataset acquired from 25 subjects. We observed that this small training dataset is sufficient for the diffusion model to generate realistic MRI-T1w images and perform well on the downstream task of deep-MRI-guided PET reconstruction. However, despite the high perceptual quality, we identified quantitative errors in the synthesized MRI images when compared to the acquired MRI-T1w images. We hypothesize that this discrepancy may be attributed to the breath of the data. Our future work involves applying this study to a dataset acquired from a larger number of subjects, with the potential to mitigate such quantitative errors in MRI synthesis.

The work presented here was only explored for FDG scans, but it could be in principle extended to other radio-tracers, as long as the PET and MRI images present high correlation. A limitation of this approach is cases where PET and MRI images show different patterns or structures, such as cases with PET-only or MRI-only lesions.

\section{Conclusion}

We showed in this work a method to produce pseudo-MRI-T1w images from PET-FDG reconstructed images using a diffusion probabilistic model for MRI-guided PET image reconstruction. Resulting images were compared to reconstructed images using the acquired MRI-T1w, reaching the same image quality, and always improving the image quality achieved with OSEM. The method was shown to work successfully for different levels of PET counts down to 25\% of the counts. Improvements for the 5\% datasets were also observed, but the images were overall rated as poor.

\section*{Acknowledgments}
The authors declare the following financial interests/personal relationships which may be considered as potential competing interests with the work reported in this paper: Jorge Cabello and Carl von Gall are full-time employees, Siemens Medical Solutions USA, Inc. G\"unther Platsch is full-time employee, Siemens Healthineers AG, Germany.

\end{document}